\documentclass{article}
\usepackage{arxiv}
\usepackage[utf8]{inputenc}
\usepackage{graphicx}
\usepackage{graphics}
\usepackage{epstopdf}
\usepackage[T1]{fontenc}    
\usepackage{url}            
\usepackage{booktabs}       
\usepackage{amsfonts}       
\usepackage{nicefrac}       
\usepackage{microtype}      

\title{Shortcomings of transfer entropy and partial transfer entropy:
       Extending them to escape the curse of dimensionality}

\author{Angeliki Papana  \\
 Department of Economics\\
 University of Macedonia\\
 Egnatias 156, Thessaloniki, Greece\\
 \texttt{angeliki.papana@gmail.com, apapana@uom.edu.gr} \\
 \And
 Ariadni Papana-Dagiasis \\
 Department of Mathematics\\
 Cleveland, Ohio, USA\\
 \texttt{a.papanadagiasis@csuohio.edu} \\
 \And
 Elsa Siggiridou\\
 Department of Electrical and Computer Engineering \\ 
 Aristotle University of Thessaloniki, Greece\\
 \texttt{esingiri@ece.auth.gr} \\
}

\begin{document}

\maketitle

\begin{abstract}
Transfer entropy (TE) captures the directed relationships between two variables. Partial transfer entropy 
(PTE) accounts for the presence of all confounding variables of a multivariate system and infers only 
about direct causality. However, the computation of PTE involves high dimensional distributions and thus 
may not be robust in case of many variables. In this work, different variants of PTE are introduced, by 
building a reduced number of confounding variables based on different scenarios in terms of their 
interrelationships with the driving or response variable. Connectivity-based PTE variants and utilizing 
the random forests (RF) methodology are evaluated on synthetic time series. The empirical findings 
indicate the superiority of the suggested variants over TE and PTE, especially in case of high dimensional
systems.
\end{abstract}


\section{Introduction}
Connectivity analysis deals with the interdependence relationships between two or more variables. There
are two distinct cases of interdependence. In the first case, the variables evolve in synchrony and we are
interested in correlations. Correlation measures express the empirical association between variables while
quantifying their dependence. In the second one, a variable drives another one and they are connected with
a causal relationship.

The investigation of the dynamical causal relationships between the variables of a multivariate system is 
essential in a variety of applications, such as physiology, e.g.~to understand the links between 
functional cerebral functions and their underlying brain mechanisms \cite{Seth15,Spencer18,Qian18}, and 
finance, e.g.~to determine the possible sources of inflation and therefore decide on policies to reduce it
\cite{Di18,Balcilar18}.

Since the pioneering work by Granger \cite{Granger69}, the advance on Granger causality (GC) has grown
considerably. Transfer entropy (TE) is its information-theoretic analog \cite{Barnett09,Schindlerova11},
relying on the same idea, i.e.~ if the prediction of a time series could be improved by incorporating the 
knowledge of past information of another one, then the latter is considered to have a causal influence on 
the former. Unlike the standard GC test, the TE does not rely on any model or assumption on the nature of 
the data, while identifies both linear and nonlinear interrelationships. TE has proved to be effective in 
a variety of simulation studies and real applications 
\cite{Siggiridou15IEEE,Porfiri18,Toriumi18,Gencaga18,Lungarella07}. For a review on TE, see 
\cite{Bossomaier16}. 

Bivariate approaches, such as GC and TE, may suffer from shortcomings, such as omitted variable bias
\cite{Lutkepohl82} and thereby can lead to erroneous conclusions. The importance of using multivariate 
methods has been emphasized in a variety of works, such as in \cite{Zachariadis06,Blinowska11,Montalto14},
since they provide a better representation of real-world interactions. Direct causality measures and tests
exploit all the available information from the whole set of the observed variables of a multivariate
system and indicate only the direct causality between variables, excluding indirect causality from 
intermediate variables, such as the causality $X \rightarrow Y$ stemming from the intermediate variable 
$Z$: $X \rightarrow Z \rightarrow Y$.

The GC has been extended in the multivariate case in order to infer only direct causality \cite{Geweke84}.
In \cite{Franciotti18}, the reliability of conditional GC (CGC) is investigated in the time domain, 
suggesting that a large number of time points decreases the reliability of the results and increases the 
number of type I errors. A high sensitivity is established, when the underlying structure of the examined 
system has the delays between the interacting nodes much larger than the
frequency resolution, the connection strength is higher than a specific percentage of the amplitude of the
driver signal and the signal to noise ratio is not very high. Finally, indirect links were revealed by the
analysis for specific connection strengths and lags. The TE has also been extended in the multivariate
case, namely the Partial TE (PTE), while different estimators of PTE have been proposed, such as based on
binning \cite{Verdes05}, correlation sums \cite{Vakorin09} and $k$-nearest neighbors (KNN)
\cite{Papana12}. The PTE though seems to be only effective for low dimensional systems
\cite{Vakorin09,Kugiumtzis13PTERV,Papana12,Papana17}.

A non-uniform embedding scheme (NUE) for the estimation of a direct coupling information measure has been 
suggested in \cite{Kugiumtzis13NUE}. However, as the dimensionality of a system increases, even the most 
effective measures utilizing dimension reduction techniques, may fail. That is because the curse of 
dimensionality re-emerges at some point when original data are of high dimensionality since the majority 
of the proposed dimension reduction methods proceed by estimating probability distributions of increasing 
dimensions. 

Further extending CGC and taking into consideration the above computational limitations, the partially 
conditioned GC (PCGC) is introduced \cite{Marinazzo12}. In this work, a subset of conditioning variables 
is used for the estimation of multivariate GC based on a mutual information criteria. The PCGC is shown to
significantly improve the performance of standard linear GC. As stated in \cite{Marinazzo12}: 
"conditioning on a small number of variables, chosen as the most informative ones for the driver node, 
leads to results very close to those obtained with a fully multivariate analysis and even better in the 
presence of a small number of samples."

In view of all the above, we suggest an ensemble of different variations of the original PTE, aiming to 
improve the performance of PTE for high dimensional data by reducing the number of the conditioning 
variables. The PTE variants can be easily implemented and are feasible on arbitrary data dimension. To
formulate the PTE variants, the connectivity relationships between the driving or response variable to the
confounding ones is examined, while also a more sophisticated method is utilized for variable selection,
namely the random forests (RF) method \cite{Ho98}. The proposed PTE variants are compared to each other,
to the bivariate TE and the multivariate PTE on known simulation systems. Results suggest that when
dimensionality increases, the PTE variants outperform both TE and PTE. Through the simulation study, the 
advantages and shortcomings of TE and PTE are displayed in detail.

The rest of the paper is organized as follows. In Section~\ref{sec:Methods}, the PTE variants are 
introduced. In Section~\ref{sec:NumericalStudies}, the performance of the PTE variants is displayed based 
on Monte Carlo simulations on known coupled systems and compared with TE and PTE. Finally, 
Section~\ref{sec:Conclusion} concludes.

\section{Methods}
\label{sec:Methods}

In this section, we briefly review the causality measures transfer entropy (TE) and partial transfer 
entropy (PTE), illustrate the significance test for the non-causality hypothesis and finally introduce the
ensemble of PTE variants.

\subsection{Transfer Entropy}

The transfer entropy (TE) is a nonlinear measure that quantifies the amount of information explained in 
$Y$ at $h$ time steps ahead from the state of $X$ accounting for the concurrent state of $Y$ 
\cite{Schreiber00}. Let ${x_t}$, ${y_t}$ be two stationary time series and
$\textbf{x}_t = [x_t,x_{t-1},\ldots,x_{(m-1)\tau}]'$ and $\textbf{y}_t 
=[y_t,y_{t-1},\ldots,y_{(m-1)\tau}]'$ the reconstructed vectors of the state space of each system, where 
$\tau$ is the delay time and $m$ is the embedding dimension. The TE from $X$ to $Y$ can be defined based 
on entropy terms as
\begin{equation}
 \mbox{TE}_{X \rightarrow Y} =                                            
- H(y_{t+h}|\textbf{x}_t,\textbf{y}_t) + H(y_{t+h}|\textbf{y}_t)         
\label{eq:TE}
\end{equation}
where $H(x)$ is the Shannon entropy of the variable $X$.

\subsection{Partial Transfer Entropy}

The partial transfer entropy (PTE) is the extension of the TE accounting for the direct causal effect of 
$X$ to $Y$ conditioning
on all the remaining variables of a multivariate system, let us denote them $Z=Z_1,\ldots,Z_{K-2}$
(confounding variables). Let us denote by $\textbf{z}_t$ the reconstructed vectors of all the variables in
$Z$. The direct causal effect of $X$ to $Y$ measured by PTE is expressed as
\begin{equation}
 \mbox{PTE}_{X \rightarrow Y|Z} =                          %
-H(y_{t+h}|\textbf{x}_t,\textbf{y}_t,\textbf{z}_t) +      %
H(y_{t+h}|\textbf{y}_t,\textbf{z}_t)                      %
 \label{eq:PTE}
\end{equation}
The entropy terms of TE and PTE are estimated here using the $k$-nearest neighbors (KNN) method
\cite{Kraskov04,Papana12}, which is stable, not significantly affected by the choice of $k$ 
\cite{Kraskov04} and gives better estimates at moderate dimensions compared to other estimators
\cite{Papana12}.

\subsection{Statistical Significance}

In order to decide whether a small positive value of TE / PTE is significant or not, resampling methods 
are required. To assess the statistical significance of the TE / PTE, the empirical null distribution is 
constructed based on resampled time series. For this, $100$ time-shifted surrogates \cite{Quiroga02} are 
constructed, randomizing only the driving time series, while the significance level is $\alpha=0.05$. If 
the estimated value of the causality measure from the original time series is at the tail of the null 
distribution formed by the surrogates, then the null hypothesis H$_0$ of non-causality is rejected. The 
$p$-value of the one-sided test is computed by applying the correction for the empirical cumulative 
density function as given in \cite{Yu01}.

\subsection{PTE variants}

Let us suppose there is a multivariate system in $K$ variables. The estimation of any direct causality 
measures for a pair of observed variables (e.g.~for $X \rightarrow Y$) involves also the remaining $K-2$ 
observed variables (e.g.~$Z=Z_1,\ldots,Z_{K-2}$). Since the computation also includes the lagged variables
in past times, the corresponding dimensionality increases even more.

We note that the following estimation scenarios can be implemented on any multivariate causality measure,
however we will focus only on PTE since the comparisons of the different scenarios must be conducted with 
respect to a specific causality measure.

First, we introduce an ensemble of connectivity-based PTE variants. They are formed so that the original 
number of conditional variables ($K-2$) is reduced. We set the number of conditioning variables equal to a
fixed number $nc$; the optimal choice of $nc$ is also investigated. First, the criteria for forming the
subset of conditioning variables relies on a connectivity measure. Specifically, the linear correlation 
coefficient (CC), $r = cov(X,Y)$ / $(\sigma_X \sigma_Y)$, and the mutual information (MI), a general 
measure of mutual dependence between two variables from information theory, defined on entropy terms as 
$I = I(X,Y)=H(X)-H(X|Y)$ are considered. 

Regarding the connectivity-based variants, if the subset of conditioning variables satisfying each 
corresponding criterion is smaller than the desired one, then this reduced number of conditioning 
variables is utilized. If none of the confounding variables satisfy the corresponding criterion, then the 
bivariate TE is estimated instead. The statistical significance of $r$ is assessed on the statistic 
$t = r \sqrt{\frac{n-2}{1-r^2}}$, which follows the $t$-distribution with $n-2$ degrees of freedom. On the
other hand, the significance of MI is assessed non-parametrically, by randomly permuting one of the two
involved time series and the $p$-values are estimated from the one sided-test for the null that the two
variables are independent.

As an alternative, we exploit the random forests (RF) methodology \cite{Ho95,Breiman01}, in order to 
determine the most 'important' determinants for a specific target variable. RF is a sophisticated method,
introduced for handling big data, such as for hundreds or even millions of observations \cite{Genuer17}. 
It is a machine learning method based on decision trees. It is a form of nonlinear regression model where 
samples are partitioned at each node of a binary tree based on the value of one selected input feature. 
The input of each tree is sampled data from the original dataset. A subset of features  is randomly
selected from the optional features to grow the tree at each node. Each tree is grown without pruning. In
this work, the implementation of RF for variable selection relies on the tree minimal depth methodology
\cite{Ishwaran10}, a method that determines the variable importance by the position of the variables in
the decision trees and thus is only based on the decision tree structures. In this case, $nc$ is 
automatically specified by the RF method and not a' priori selected as in case of the connectivity-based
variants.

The ensemble of the proposed connectivity-based (1A - 4B) and RF-based (5A - 5C) variants of PTE consider 
different subsets of conditioning variables selected from the original set of confounding variables. 
Specifically, the subset of the conditioning variables is defined by the following criteria:  
\newline
\textbf{1A}: Choose the most correlated variables to the driving one based on $r$ \\
\textbf{1B}: Choose the most correlated variables to the driving one based on $I$ \\
\textbf{2A}: Choose the least associated variables to the driving one based on $r$ \\
\textbf{2B}: Choose the least associated variables to the driving one based on $I$ \\
\textbf{3A}: Choose the most correlated variables to the response one based on $r$ \\
\textbf{3B}: Choose the most correlated variables to the response one based on $I$ \\
\textbf{4A}: Choose the least associated variables to the response one based on $r$ \\
\textbf{4B}: Choose the least associated variables to the response one based on $I$. \\
\textbf{5A}: Choose the most associated variables to the driving one based on RF \\
\textbf{5B}: Choose the most associated variables to the response one based on RF \\
\textbf{5C}: Choose the most associated variables to the driving and response variable based on RF 

\section{Numerical studies}  
\label{sec:NumericalStudies}

The performance of the PTE variants is examined through simulation experiments. Three known coupled systems are considered including linear and nonlinear couplings. For comparison reasons, we also estimate the bivariate transfer entropy (TE) and the original multivariate transfer entropy (PTE), while the connectivity-based PTE variants are estimated by varying the number of the conditioning variables.

 We formulate 100 realizations from each simulation system, with time series lengths $n=512, 1024, 2048$ and estimate the causality measures for all directions. The TE, PTE and PTE variants are computed for $h = 1$ and $\tau =1$ (as in definition of TE \cite{Schreiber00}) and the embedding dimension $m$ is fixed based on the equations of the system. Finally, the number of neighbors for the KNN method is equal to $k = 10$ ($k$ does not affect the performance of the KNN estimator \cite{Papana13}).

The performance of the measures is evaluated in terms of binary classification tests, where 'positive' corresponds to the case of causality and 'negative' to the non-causal case. Specifically, results are discussed by considering three classification measures; sensitivity, specificity and F1-score. First, we estimate the number of 'positives' and 'negatives' found, where:
 True Positive (TP) stands for causal link correctly classified as positive, True Negative (TN) for a non-causal links correctly classified as negatives, False Positive (FP) for non-causal inks erroneously classified as positive, and False Negative (FN) for a causal link erroneously classified as negatives. The sensitivity (Sen) or true positive rate measures the proportion of actual positives that are correctly identified as such by the measure, i.e. we evaluate the ability of the measures to detect the true causal effects:
\begin{equation}
  Sensitivity  = TP /(TP+FN) \nonumber
  \label{eq:Sensitivity}
\end{equation}
The specificity (spec) or true negative rate quantifies the proportion of actual negatives that are correctly identified as such, i.e.
indicates the ability of a measure to correctly detect the non-causal variables:
\begin{equation}
  Specificity  = TN / (TN+FP) \nonumber
  \label{eq:Specificity}
\end{equation}
In theory, sensitivity and specificity are independent in the sense that it is possible that they both achieve their best value
($100\%$). Finally, the F1-score determines the overall test's accuracy:
\begin{equation}
  F1-score = 2TP /(2TP+FP+FN). \nonumber
  \label{eq:F1-score}
\end{equation}

\subsection{(S1) Simulation system 1}

The first simulation system is a VAR(4) process in five variables with $X_1 \rightarrow X_2$, $X_1 \rightarrow X_4$, $X_2
\rightarrow X_4$, $X_4 \rightarrow X_5$, $X_5 \rightarrow X_1$, $X_5 \rightarrow X_2$ and $X_5 \rightarrow X_3$ \cite{Schelter06}
\begin{eqnarray}
  x_{1,t} & = & 0.4x_{1,t-1} - 0.5x_{1,t-2} + 0.4x_{5,t-1} + \epsilon_{1,t} \nonumber\\
  x_{2,t} & = & 0.4x_{2,t-1} - 0.3x_{1,t-4} + 0.4x_{5,t-2} + \epsilon_{2,t} \nonumber\\
  x_{3,t} & = & 0.5x_{3,t-1} - 0.7x_{3,t-2} - 0.3x_{5,t-3} + \epsilon_{3,t} \nonumber\\
  x_{4,t} & = & 0.8x_{4,t-3} + 0.4x_{1,t-2} + 0.3x_{2,t-3} + \epsilon_{4,t} \nonumber\\
  x_{5,t} & = & 0.7x_{5,t-1} - 0.5x_{5,t-2} - 0.4x_{4,t-1} + \epsilon_{5,t}, \nonumber
\label{eq:Schelter}
\end{eqnarray}
where $\epsilon_{i,t}$, $i=1,\ldots,5$ are independent to each other Gaussian white noise processes with unit standard deviation.

Causality measures are estimated for $m = 4$ (the maximum delay in the system's equations). First, results for TE and PTE are reported in Table~\ref{tab:S1_TE_PTE}. Sensitivity (Sen), specificity (Spec) and F1-score are computed from 100 realization of the system for each time series length $n$ and finally as a mean over all $n$. Bivariate TE has a higher sensitivity compared to PTE, while multivariate PTE outperforms TE regarding specificity. In particular, TE has a high sensitivity which increases with $n$, but has a relatively low specificity that decreases as $n$ gets larger. On the other hand, PTE may has a slightly lower sensitivity compared to TE, but it increases  with $n$. Additionally, specificity for PTE decreases with $n$. Overall, the PTE ($88.24\%$) outstands TE ($75.45\%$), based on the mean F1-score over all $n$.
\begin{table}[ht]
\centering
 \caption{Classification measures for TE and PTE for S1, as a mean over all realizations for each time length $n$.}
  {\begin{tabular}{c ccc   ccc} \hline
           &\multicolumn{3}{c}{TE}    & \multicolumn{3}{c}{PTE}         \\ \hline
                    & Sen     & Spec   & F1-score        & Sen    & Spec    & F1-score  \\
 $n=512$    & 98.43  & 75       & 80.89  & 79.86 & 92.92  & 82.72   \\
 $n=1024$  & 99.86  & 64.69  & 75.58  & 93.71 & 92.77  & 90.72   \\
 $n=2048$  & 100     & 53.15  & 69.87  & 98.29 & 90.38  & 91.27   \\  \hline
  MEAN     & 99.43  & 64.28  & 75.45  & 90.62 & 92.03   & 88.24     \\   \hline
 \end{tabular}
 \label{tab:S1_TE_PTE}}
\end{table}

\begin{figure}[ht]
\centering
 \includegraphics[scale=.36]{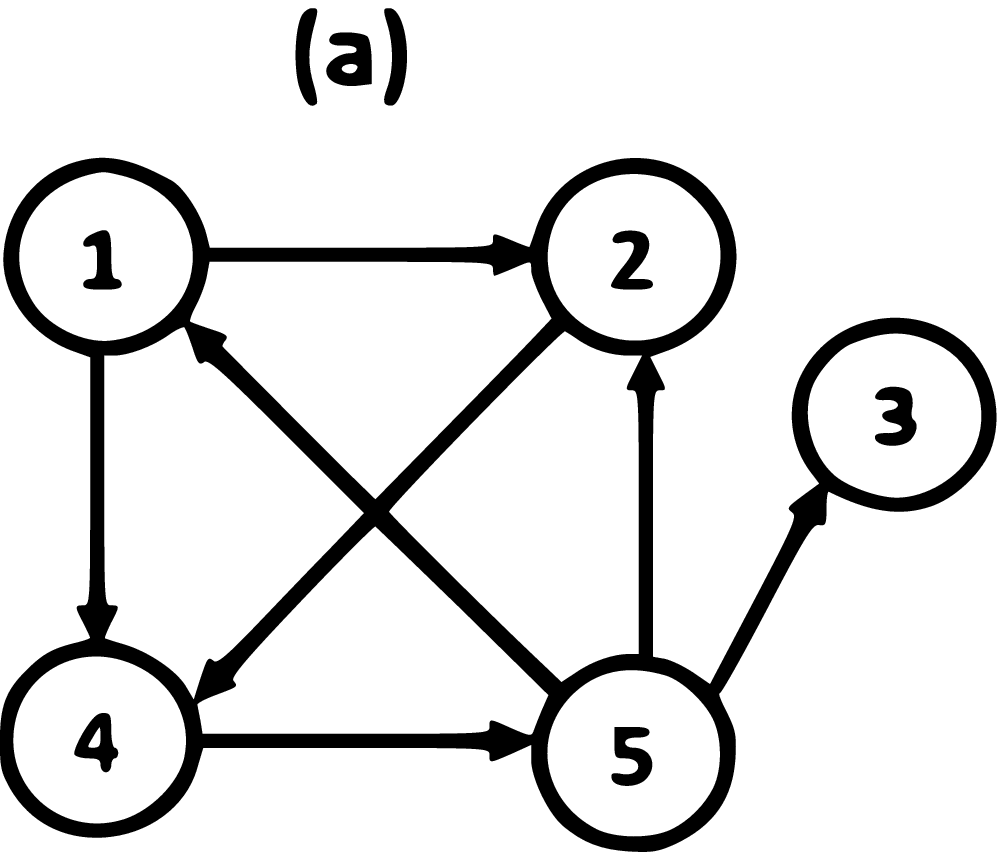}
 \includegraphics[scale=.36]{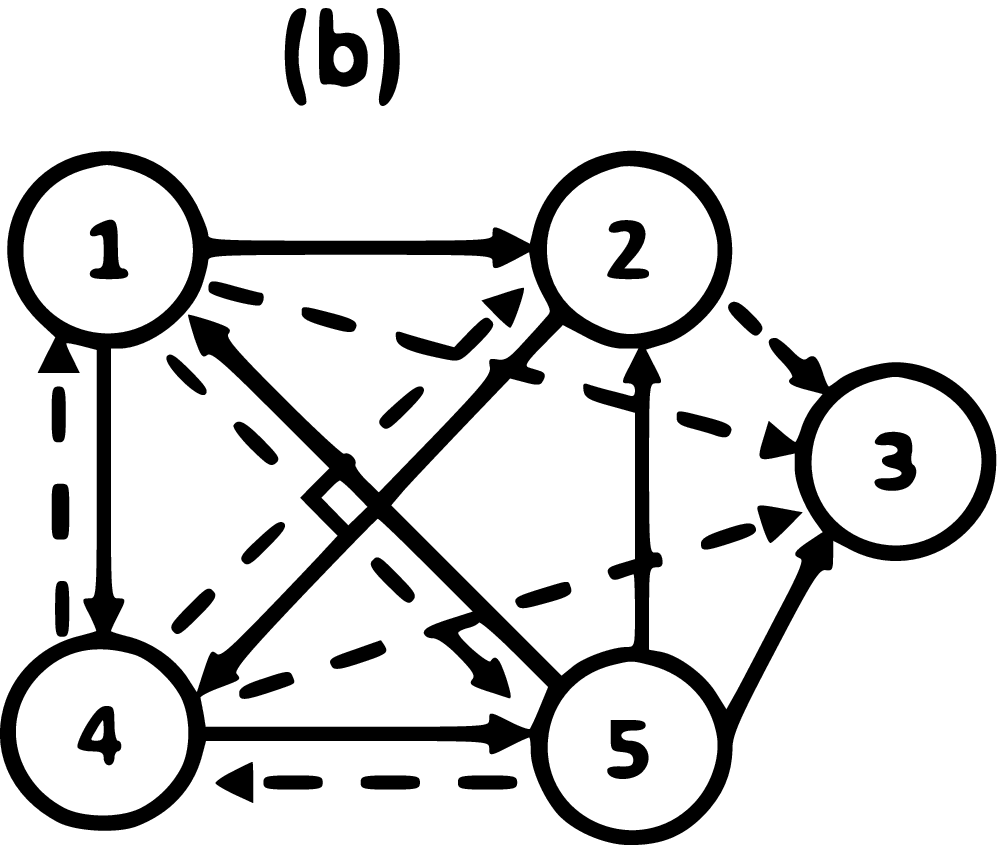}
 \includegraphics[scale=.36]{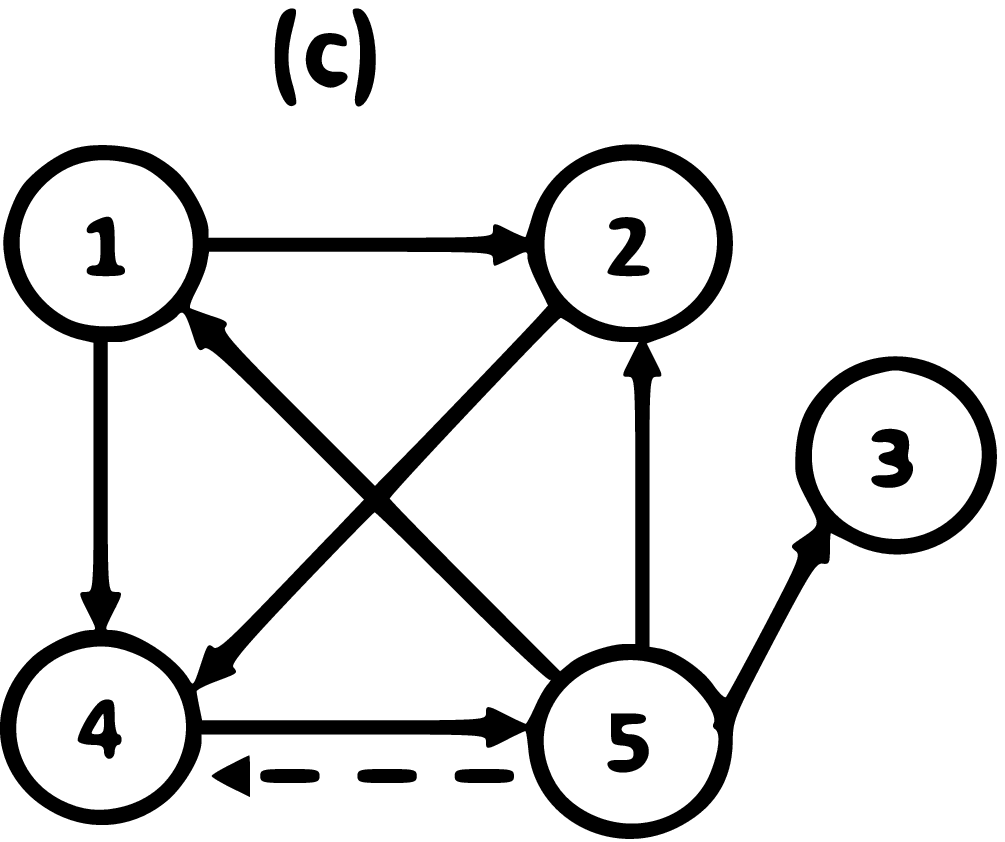}
 \caption{The true network of S1 in (a) and the extracted networks based on TE in (b)
  and on PTE in (c). Dashed lines indicate indirect causalities.}
 \label{fig:Syst1Network}
\end{figure}

In order to further analyze the above quantitative results, the significant causal effects over all realizations are examined for all couples of variables. As expected, TE indicates both direct and indirect causal effects. As $n$ increases, the percentages of significant indirect causalities also increase, leading to a decreasing specificity and thus a decreasing overall performance (F1-score). Since PTE is multivariate, is expected to only indicate direct couplings, while for $K=5$, no major estimation
problems should arise due to high dimensionality. For small $n$, the percentages of significant causality are low only for $X_2 \rightarrow X_4$ ($28\%$ for $n=512$, $68\%$ for $n=1024$, $88\%$ for $n=2048$) and $X_5 \rightarrow X_3$ ($64\%$ for $n=512$, $89\%$ for $n=1024$, $100\%$ for $n=2048$), while the indirect coupling $X_5 \rightarrow X_4$ is obtained ($24\%$ for $n=512$, $31\%$ for $n=1024$, $50\%$ for $n=2048$). The above findings can be summarized in Fig.~\ref{fig:Syst1Network}, where the true network and the corresponding detected networks by TE and PTE are presented.

The PTE variants that rely on the connectivity criteria are estimated by setting the number of conditioning variables equal to $nc=1$ and $nc=2$. Empirical findings from S1 show that CC gives many significant linear correlations which increase with $n$. On the other hand, MI detects much fewer 
correlations compared to linear CC. All the PTE variants have slightly higher sensitivity compared to PTE but lower compared to TE (Table~\ref{tab:SchelterPTEvariants}). Further, they present higher specificity compared to TE but lower specificity compared to PTE. As $n$ increases, the sensitivity of the variants slightly increases while the specificity decreases. Their overall performance worsens since specificity mostly affects it.  
\begin{table}[ht]
\centering
  \caption{Classification measures for connectivity-based PTE variants for S1, as a mean over all realizations for each time length $n$ and number of conditioning variables $nc$.}
 {\begin{tabular}{c  ccc  ccc  ccc} \hline
     &\multicolumn{9}{c}{$nc=1$}            \\    \hline
     &\multicolumn{3}{c}{$n=512$}   & \multicolumn{3}{c}{$n=1024$}   & \multicolumn{3}{c}{$n=2048$}   \\ 
     & Sen & Spec   &  F1-score & Sen   & Spec  &  F1-score  & Sen  & Spec    &  F1-score   \\  \hline
 1A &  92.57 & 85.85 & 84.78     & 98.86  & 80.23 & 84.45     & 100       & 73.38    & 80.55 \\  
 1B &  94.43 & 81.61 & 82.97     & 99.29  & 75       & 81.32     & 100      & 67.62    & 77.31 \\  
 2A & 91.86  & 84.92 & 83.99     & 96.86  & 80.23  & 83.41     & 99.86   & 70.23    & 78.83  \\ 
 2B &  91    & 85.62 & 83.96     & 96.29  & 78.15  & 81.91     & 99.71  & 70.92    & 79.12  \\  
 3A &  93     & 85.54 & 84.88     & 97.71  & 78.77  & 82.79     & 99.86  & 72.38    & 79.98  \\ 
 3B &  95.14  & 79.31 & 81.95     & 98.71  & 72.62  & 79.46     & 100     & 67.46     & 77.21  \\  
 4A &  90.14  & 85.54 & 74.29     & 96.14  & 79.92  & 82.75     & 99.86  & 64         & 75.4   \\ 
 4B &  90.71  & 85.31 & 83.74     & 97.43  & 78        & 82.26    & 99.57  & 70.92    & 78.97  \\  \hline
            &\multicolumn{9}{c}{$nc=2$}                                                   \\    \hline
 1A & 90.14  & 85.77  & 83.45     & 95.86  & 83.23   & 84.86    & 98.71  & 76.92  & 82.2   \\  
 1B & 94       & 81.85  & 82.9	& 99.29  & 75.85   & 81.84    & 99.71  & 69.46   & 78.18  \\
 2A & 88.43 & 87.85  & 84.03	& 96.14  & 83.85   & 85.42    & 99.86  & 70.23   & 78.83  \\
 2B & 85.71 & 88.92  & 83.32	& 95.57  & 85.77   & 86.4      & 99       & 80.38   & 84.51  \\
 3A & 91.43 & 86.85  & 84.92	& 96.43  & 83.92   & 85.7      & 99.86  & 81.92   & 86.09  \\
 3B & 95      & 79.46  & 81.95	& 98.43  & 73.23   & 79.72    & 100     & 68.46   & 77.78  \\
 4A & 87.14 & 88       & 83.53	& 96.29  & 80.23   & 83.04    & 99.57  & 64.38   & 75.48  \\
 4B & 85.57 & 90.38  & 84.19	& 86.77  & 86.77   & 86.86    & 98.14  & 81        & 84.62  \\  \hline
\end{tabular}
\label{tab:SchelterPTEvariants}}
\end{table}

The number of conditioning variables also influences the performance of the connectivity-based variants. For $nc = 1$, we have always slightly higher sensitivity compared to $nc =2$. On the contrary, for $nc = 2$, we have always slightly higher specificity compared to $nc =1$. The best performance among 
the PTE variants is obtained for variant 3A, i.e. when choosing the conditioning variables as the most correlated to the response one. The mean F1-score over all realizations and time series lengths for 3A is $85.57\%$, very close to PTE's mean performance ($88.24\%$), but much higher than TE's ($75.45\%$).

Regarding the RF-based variants, 5C coincides with the multivariate case, i.e. coincides with PTE, thus scoring the highest F1-score. The variant 5A slightly outperforms all the considered connectivity-based variants, achieving a mean F1-score equal to $86.58\%$ and 5B follows, scoring very closely (mean F1-score $86.44\%$) (Table~\ref{tab:SchelterPTEvariantsRF}). 
\begin{table}[ht]
\centering
\caption{Classification measures for RF-based PTE variants for S1, as a mean over all realizations for each time length $n$.}
 {\begin{tabular}{c  ccc  ccc  ccc} \hline
      &\multicolumn{3}{c}{$n=512$}   & \multicolumn{3}{c}{$n=1024$}   & \multicolumn{3}{c}{$n=2048$}   \\ 
      & Sen & Spec   &  F1-score  & Sen  & Spec  &  F1-score  & Sen  & Spec    &  F1-score \\  \hline
 5A &  91.71 & 85.23  &  83.93 & 95   & 87.15  & 87.12   &  98.71   & 86.31  & 88.69           \\   
 5B &  91.57  & 83.85 &  82.97 & 95.14  & 87.69  & 87.63   &  98.57   & 86.5  & 88.72  \\  \hline
\end{tabular}
\label{tab:SchelterPTEvariantsRF}}
\end{table}


\subsection{(S2) Simulation system 2}

We consider a system with linear ($X_1 \rightarrow X_3$, $X_4 \leftrightarrow X_5$) and nonlinear couplings ($X_1 \rightarrow X_2$,  
$X_1 \rightarrow X_4$) given by the equations:
\begin{eqnarray}
  x_{1,t} & = & 0.95 \sqrt{2} x_{1,t-1} - 0.9025 x_{1,t-2} + \epsilon_{1,t} \nonumber\\
  x_{2,t} & = & 0.5 x_{1,t-2}^2 + \epsilon_{2,t}  \nonumber\\
  x_{3,t} & = & -0.4x_{1,t-3} + \epsilon_{3,t} \nonumber\\
  x_{4,t} & = & -0.5 x_{1,t-2}^2 + 0.25 \sqrt{2} x_{4,t-1} + 0.25 \sqrt{2} x_{5,t-1} + \epsilon_{4,t}, \nonumber\\
  x_{5,t} & = & -0.25 \sqrt{2} x_{4,t-1} + 0.25 \sqrt{2} x_{5,t-1} + \epsilon_{5,t}, \nonumber
\label{eq:Montalto}
\end{eqnarray}
with Gaussian noise terms as in S1 \cite{Montalto14}.

The TE indicates correctly the true causality. However, it also detects the indirect causal link $X_1 \rightarrow X_5$, e.g. for $n=2048$, with percentage $100\%$. Additionally, the spurious links $X_2 \rightarrow X_3$ ($100\%$), $X_2 \rightarrow X_4$ ($100\%$), $X_2 \rightarrow X_5$ ($100\%$), $X_3 \rightarrow X_2$ ($80\%$), $X_3 \rightarrow X_4$ ($78\%$), $X_3 \rightarrow X_5$ ($100\%$), $X_4 \rightarrow X_2$ ($100\%$), $X_4 \rightarrow X_3$ ($100\%$), $X_5 \rightarrow X_2$ ($100\%$), $X_5 \rightarrow X_3$ ($100\%$) are obtained (percentages are indicatively shown for $n=2048$).

The PTE indicates almost all the true links; it fails to identify the true relationship $X_5 \rightarrow X_4$. Further, it indicates the indirect effect $X_1 \rightarrow X_5$ ($100\%$ for $n=2048$) and the false ones $X_2 \rightarrow X_4$ ($100\%$), $X_2 \rightarrow X_5$ ($100\%$), $X_3 \rightarrow X_5$ ($21\%$), $X_5 \rightarrow X_2$ ($18\%$). 

The extracted outcomes of the binary classification measures regarding TE and PTE are displayed on Table~\ref{tab:MontaltoTE_PTE}.
The overall performance of TE is low. TE has a high sensitivity that sligthly increases with $n$, however obtains a very low specificity which decreases with $n$. On the other hand, PTE has lower sensitivity compared to TE but still on a high level. It also achieves a high specificity. Both the sensitivity and specificity of PTE decrease with $n$. The mean F1-score of TE and PTE over all $n$ is $39.12\%$ and $65.7\%$, respectively. Overally, the PTE outperforms the TE, however cannot succeed a high mean F1-score. 
\begin{table}[ht]
\centering
  \caption{Classification measures for TE and PTE for S2, as a mean over all realizations for each $n$.}
 {\begin{tabular}{c ccc   ccc} \hline
           &\multicolumn{3}{c}{TE}    & \multicolumn{3}{c}{PTE}         \\ \hline
                    & Sen     & Spec   & F1-score       & Sen      & Spec    & F1-score       \\
 $n=512$    & 89.80  & 33.33  & 46.38   & 82.4    & 80.27  & 68.6   \\
 $n=1024$  & 93.20  & 29.93  & 46.27   & 81.6    & 76.27  & 64.8   \\
 $n=2048$  & 96.80  & 29.47  & 47.47    & 80.6   & 75.33  & 63.58  \\  \hline
  MEAN        & 93.27   & 30.91  & 39.12   & 81.53  & 78.35  & 65.7     \\   \hline
 \end{tabular}
\label{tab:MontaltoTE_PTE}}
\end{table}

For S2, the MI detects more correlations compared to the linear CC. Cumulative results over all realizations and time series lengths are presented in Table~\ref{tab:S2_PTEvariants}, for the connectivity based PTE variants. Sensitivity is high and increases with $n$ for both $nc$ (varies from $82.2\%$ to $96.4\%$), specificity is low (varies from $34.53\%$ to $72\%$) but increases with $n$. Both sensitivity and specificity are in-between the estimated corresponding values from TE and PTE. The PTE variants have slightly higher sensitivity compared to PTE ($80.60\%$) but lower compared to TE ($96.80\%$). The PTE variants have higher specificity compared to TE ($30.91\%$) but lower specificity than PTE ($75.33\%$). All variants outperform TE, however cannot reach the mean F1-score of PTE due to their low specificity. For $nc=1$, we have equal or slightly higher sensitivity compared to $nc=2$. For $nc=2$, we have always slightly higher specificity compared to $nc=1$. The best performance among PTE variants is obtained for the case 1B (mean F1-score $61.15\%$), i.e. when choosing the conditioning variables as the most correlated to the driving one using MI. 
\begin{table}[ht]
\centering
\caption{Classification measures for connectivity-based PTE variants for S2, as a mean over all realizations and time series lengths for each $nc$.} 
 {\begin{tabular}{c  ccc  ccc} \hline
            &\multicolumn{3}{c}{$nc=1$}   &\multicolumn{3}{c}{$nc=2$}   \\    \hline
      & Sen      & Spec    &  F1-score        & Sen     & Spec    &  F1-score              \\  \hline
 1A &  84.33 & 55.33   & 53.3      & 84.93  & 60.56  & 56.39         \\  
 1B &  89.93 & 54.76   & 55.51     & 85.73  & 67.91  & 61.15       \\  
 2A & 90.27  & 42.8     & 50.31     & 86.4   & 45.4     & 49.85       \\ 
 2B &  92.87  & 38.49  & 49.55     & 92.87  & 39.53  & 50             \\  
 3A &  84.33 & 47.16   & 49.78     & 84.53  & 49.24  & 50.74       \\ 
 3B &  84.13  & 48.6    & 50.33     & 83.53  & 49.98  & 50.72       \\  
 4A &  90.33  & 48.53  & 52.66     & 85.67  & 55.53  & 54.03       \\ 
 4B &  91.07  & 47.71  & 52.61     & 90.73  & 48.58  & 52.89         \\   \hline
\end{tabular}
\label{tab:S2_PTEvariants}}
\end{table}

Their poor performance is due to the detection of both indirect and spurious causalities. For case 1B, the coupling $X_5 \rightarrow X_4$ 
is detected with a low percentage over the 100 realizations, e.g. for $n=2048$ and $nc=2$, the percentage of significant causality is $29\%$, the indirect causality $X_1 \rightarrow X_5$ with $100\%$, and finally the spurious couplings $X_2 \rightarrow X_3$ ($97\%$), $X_2 \rightarrow X_4$ ($100\%$),  $X_2 \rightarrow X_5$ ($90\%$), $X_4 \rightarrow X_2$ ($54\%$), $X_3 \rightarrow X_5$ ($28\%$) and $X_4 \rightarrow X_3$ ($26\%$) are also indicated.

Concerning the RF-based PTE variants, 5C again coincides with PTE, achieving the highest F1-score. The mean F1-score of 5A and 5B is $58.38\%$ and $50.27\%$, respectively (Table~\ref{tab:MontaltoPTEvariantsRF}). Concerning S2, the connectivity-based PTE variants outstand cases 5A and 5B. Although sensitivity is high for RF-based variants, it is at a lower level compared to the connectivity-based ones. Their low specificity affects their overall performance. 
\begin{table}[ht]
\centering
  \caption{Classification measures for RF-based PTE variants for S2, as a mean over all realizations for each $n$.}
 {\begin{tabular}{c  ccc  ccc  ccc} \hline
      &\multicolumn{3}{c}{$n=512$}   & \multicolumn{3}{c}{$n=1024$}   & \multicolumn{3}{c}{$n=2048$}   \\ 
      & Sen & Spec & F1-score & Sen  & Spec  &  F1-score   & Sen & Spec &  F1-score  \\  \hline
 5A & 87.4   & 64.4   & 59.88 & 87.6    & 62.07   & 58.23  & 85.2   & 61.8  & 57.04  \\   
 5B & 82.4   & 52.67  & 51.61 & 83  & 47.87  & 49.22 & 83.8    & 48.87   & 49.97  \\ \hline
\end{tabular}
\label{tab:MontaltoPTEvariantsRF}}
\end{table}


\subsection{(S3) Simulation system 3}

We consider a nonlinear dynamical system, the coupled Henon maps (discrete time) given by the equations:
\begin{eqnarray}
 x_{i,t} & = & 1.4 - x_{i,t-1}^2 + 0.3x_{i,t-2}, i=1,4 \nonumber\\
 x_{i,t} & = & 1.4 - \left(0.5c(x_{i-1,t-1}+x_{i+1,t-1})+(1-c)x_{i,t-1}\right)^2
 + 0.3x_{i,t-2}, i=2,\ldots,18 \nonumber
 \label{eq:s1}
\end{eqnarray}
\cite{Kugiumtzis13NUE}. The system can be defined for an arbitrary number of variables $K$. The coupling strength is equal to $c=0.2$ (intermediate coupling strength). Estimations are displayed for the embedding dimension $m =2$ (selected based on the equations of the system).

First, we set $K=5$. For S3, MI detects only few more correlations compared to the linear CC. The binary classification measures concerning TE and PTE 
are presented on Table~\ref{tab:cHenonK5}. Both TE and PTE correctly find the true connections of the system. The TE achieves a high sensitivity and a slightly lower specificity, resulting in a decreasing F1-score that varies from $84.7\%$ to $75.34\%$. It turns out that the decreased performance of TE is mainly due to the detection of indirect causal effects, which is expected as TE is bivariate. For example, for $n=2048$, TE indicates $X_1 \rightarrow X_3$ ($64\%$), $X_2 \rightarrow X_4$ ($100\%$),  $X_4 \rightarrow X_2$ ($100\%$), $X_5 \rightarrow X_3$ ($69\%$). However, the spurious causal effect $X_2 \rightarrow X_1$ ($18\%$) is also obtained. On the other hand, PTE does not indicate indirect links but falsely shows $X_2 \rightarrow X_1$ ($24\%$) and $X_4 \rightarrow X_5$ ($23\%$). The mean F1-score of TE and PTE over all realizations and $n$ is $80.37\%$ and $75.45\%$, respectively.
\begin{table}[ht]
\centering
\caption{Classification measures for TE and PTE for S3 in $K=5$ variables, as a mean over all realizations for each $n$.}
 {\begin{tabular}{c ccc   ccc} \hline
           &\multicolumn{3}{c}{TE}    & \multicolumn{3}{c}{PTE}           \\ \hline
                 & Sen     & Spec   & F1-score       & Sen      & Spec    & F1-score       \\
 $n=512$    & 99.17  & 84.29  & 84.7     & 44.67   & 94.07  & 52.32   \\
 $n=1024$  & 100     & 79.36  & 81.07   & 83.17   & 93.03  & 82.79   \\
 $n=2048$  & 100     & 71.36  & 75.34   & 97.83   & 92.43  & 91.23   \\  \hline
 MEAN       & 99.72  & 78.34  & 80.37   & 75.22   & 93.19  & 75.45   \\   \hline
 \end{tabular}
\label{tab:cHenonK5}}
\end{table}

The connectivity-based PTE variants have slightly lower sensitivity compared to TE but much higher compared to PTE (Table~\ref{tab:S3PTEvariants}). Conclusively, they outperform TE and PTE. Specifically, the mean estimated F1-score over all realizations and time series for the TE is $80.37\%$, for the PTE is $75.44\%$ and for the PTE variants varies from $82.63\%$ to $89.06\%$. They have slightly lower sensitivity compared to TE but much higher compared to PTE. Their specificity is larger compared to TE but obtain lower specificity compared to PTE. For $nc=1$, we have always slightly higher sensitivity compared to $nc=2$. For $nc=2$, we have equal or slightly higher specificity compared to $nc=1$. The best performance among PTE variants is obtained for the cases 3B, i.e. when choosing the conditioning variables as the most correlated to the response one.
\begin{table}[ht]
\centering
  \caption{Classification measures for connectivity-based PTE variants for S3 in $K=5$ variables, as a mean over all realizations and time series lengths for each $nc$.}
 {\begin{tabular}{c  ccc  ccc} \hline
            &\multicolumn{3}{c}{$nc=1$}   &\multicolumn{3}{c}{$nc=2$} \\    \hline
      & Sen     & Spec   &  F1-score  & Sen     & Spec     &  F1-score                \\  \hline
 1A & 97.56  & 86.86  & 86.12       & 96.17    & 86.62   & 85.22         \\  
 1B & 96.89  & 89.07  & 87.65       & 94.83    & 89        & 86.3         \\  
 2A & 96.17  & 85.07  & 83.86       & 91.72    & 87.62   & 83.28       \\ 
 2B & 96.33  & 84.17  & 83.32       & 92.78    & 85.95   & 82.63      \\  
 3A & 97.61  & 89.64  & 88.54       & 95.94    & 90.74   & 88.58       \\ 
 3B & 97.67  & 90.14  & 89.06       & 95.11    & 91.14   & 88.44       \\  
 4A & 95.94  & 84.98  & 83.78       & 91.89    & 87.02   & 83.12       \\ 
 4B & 95.94  & 86.07  & 84.57       & 93.33    & 86.14   & 83.14       \\   \hline
\end{tabular}
\label{tab:S3PTEvariants}}
\end{table}

\begin{figure}[ht]
\centering
 \includegraphics[scale=.40]{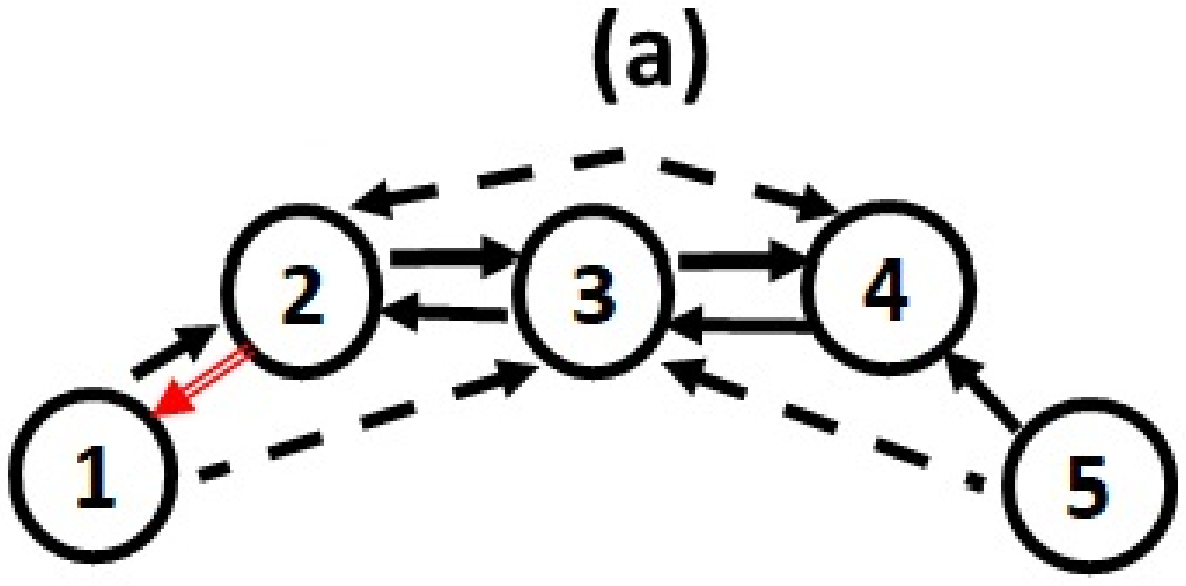}
 \includegraphics[scale=.40]{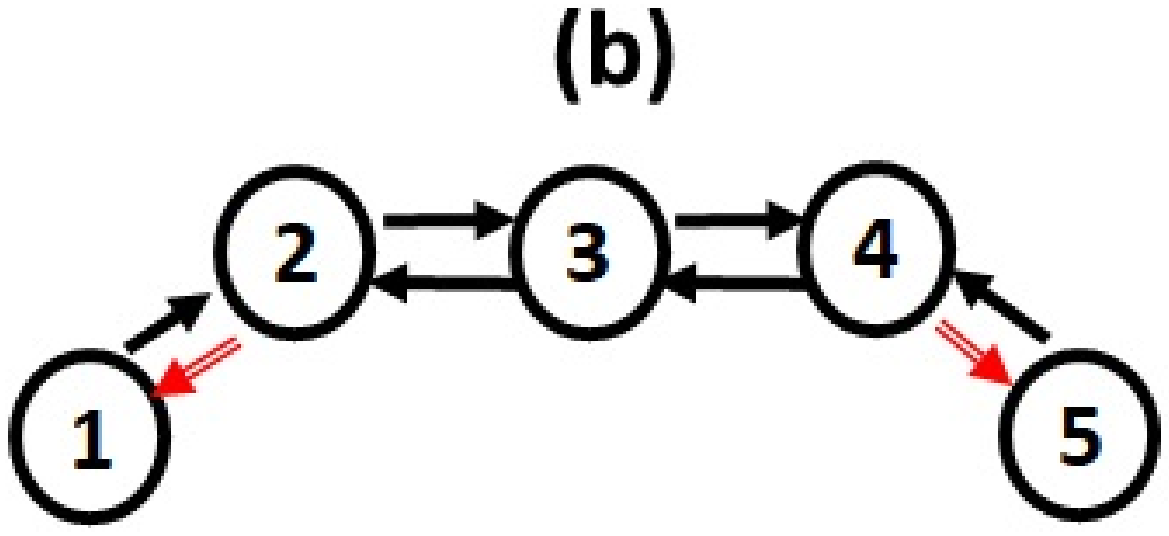}
 \includegraphics[scale=.40]{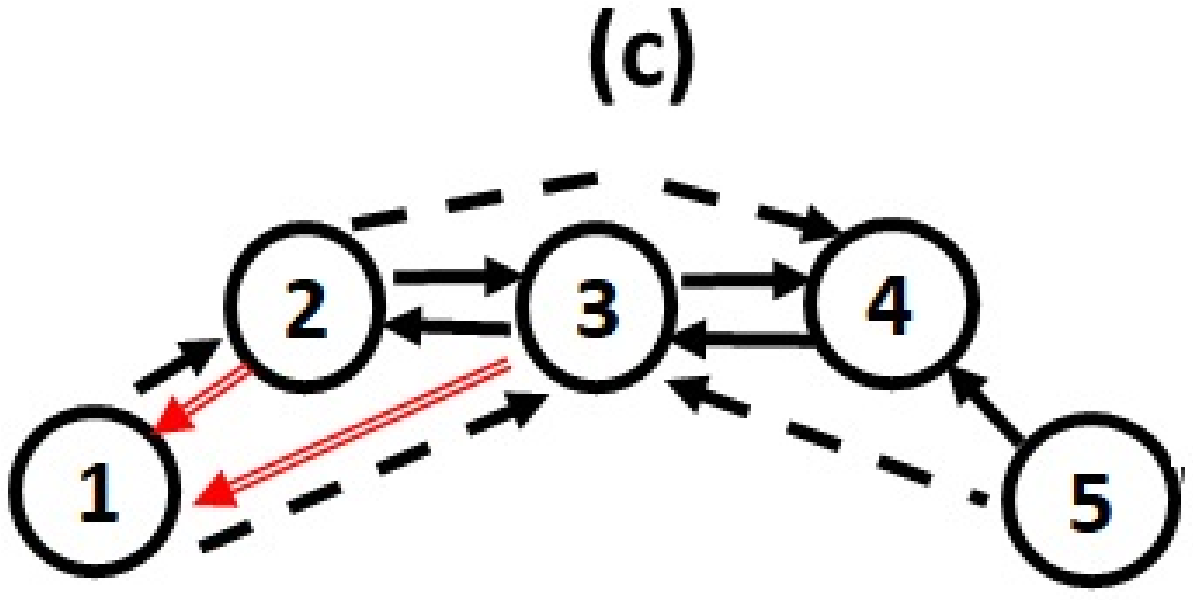}
 \caption{For S3 in $K=5$ variables ($c=0.2$), the extracted networks based on TE in (a), on PTE in (b) and on PTE variant 3B with $nc=1$ in (c). 
Dashed lines indicate indirect causalities. Double lines indicate spurious links.}
 \label{fig:Syst3Network}
\end{figure}

In order to further demonstrate the performance of the optimal connectivity-based PTE variant, namely case 3B for $nc=2$, we display the estimated connectivity network from one realization of S3 (K=5,c=0.2) with $n=2048$, along with the corresponding networks for TE and PTE (Fig.~ \ref{fig:Syst3Network}). We further note the respective percentages of rejection of the non-causality hypothesis over the 100 realizations for each one of the detected reported causal links. All the true couplings are detected by 3B. The indirect causal links $X_1 \rightarrow X_3$ ($27\%$), $X_2 \rightarrow X_4$ ($14\%$), $X_5 \rightarrow X_3$ ($24\%$) and the spurious links $X_2 \rightarrow X_1$ ($25\%$) and $X_3 \rightarrow X_1$ ($12\%$) are also obtained.

Similarly to connectivity-based variants, the sensitivity of the RF-based variants increases with $n$, while the specificity decreases. The optimal number of conditioning variables is not significantly affected by the time series length and gets small values. The most significant variables exported from RF seem to coincide with the variables that are truly coupled, i.e. for one realization of the system, we get the following outcome: the most informative variables for variable 1 is variable 2, the most informative variables for variable 2 are variables 3 and 4, the most informative variables for variable 3 are variables 2 and 4, etc. The RF-based variant 5B, achieves the highest total mean F1-score over all realizations and time series lengths ($90.36\%$), capturing all the causal influences and slightly outperforming case 3B (Table~\ref{tab:S3K5PTEvariantsRF}). Closely scores variant 5A, with overall mean F1-score $88.67\%$, while 5C again coincides with PTE and therefore has the worst performance for S3 in $K=5$ variables.
\begin{table}[ht]
\centering
  \caption{Classification measures for RF-based PTE variants for S3 in $K=5$ variables, as a mean over all realizations for each $n$.}
 {\begin{tabular}{c  ccc  ccc  ccc} \hline
      &\multicolumn{3}{c}{$n=512$}   & \multicolumn{3}{c}{$n=1024$}   & \multicolumn{3}{c}{$n=2048$}   \\ 
      & Sen  & Spec    &  F1-score  & Sen & Spec  &  F1-score   & Sen   & Spec &  F1-score  \\  \hline
 5A & 92.33  & 92.64   & 88.33 & 99.33   & 90.14  &  86.16 & 100  & 92.64   & 83.68      \\   
 5B & 95.83   & 93.36   & 90.9 & 99.33   & 91.71    & 91.33 &  99.33  & 91.71   & 87.95   \\ \hline
\end{tabular}
\label{tab:S3K5PTEvariantsRF}}
\end{table}

Finally, we evaluated all measures for S3 ($c=0.2$) in $K=10$ variables. Similar results are reported from S3 in $K=10$ as for $K=5$ (Table~\ref{tab:cHenonK10}). The TE indicates all the true couplings and 
therefore has high sensitivity. But its specificity is low because many indirect causal effects and spurious ones are captured. For example, for $n=2048$, TE falsely finds the links $X_2 \rightarrow X_1$ ($13\%$) and $X_{10} \rightarrow X_9$ ($25\%$). On the other hand, the PTE has high specificity but very low sensitivity. Although PTE also detects the true couplings, requires large time series lengths to achieve high percentages of declining the non-causal null hypothesis. Indicatively for $n=2048$, some links along with the extracted percentage over all the realizations are the following: 
$X_1 \rightarrow X_2$ ($79\%$), $X_2 \rightarrow X_3$ ($44\%$), $X_3 \rightarrow X_4$ ($15\%$), $X_4 \rightarrow X_5$ ($16\%$), $X_3 \rightarrow X_2$ ($37\%$). 
The TE outperforms the PTE.
\begin{table}[ht]
\centering
  \caption{Classification measures for TE and PTE for S3 in $K=10$ variables, as a mean over all realizations for each $n$.}
 {\begin{tabular}{c ccc   ccc} \hline
           &\multicolumn{3}{c}{TE}    & \multicolumn{3}{c}{PTE}           \\ \hline
                 & Sen     & Spec   & F1-score       & Sen      & Spec    & F1-score        \\
 $n=512$    & 96.81  & 63.2   & 53.16    & 23.5     & 95.47  & 31.6   \\
 $n=1024$  & 99.81  & 52.19  & 40.32    & 28.38   & 94.04  & 35.38   \\
 $n=2048$  & 100     & 44.35  & 35.28   & 30        & 93.27  & 36.18   \\   \hline
  MEAN       & 98.88  & 53.25  & 48.2     & 27.29   & 94.26  & 34.39   \\   \hline
 \end{tabular}
\label{tab:cHenonK10}}
\end{table}

Concerning the connectivity-based PTE variants, the number of conditioning variables varies from $2$ to $7$. For increasing $nc$, sensitivity decreases, while specificity increases. The best performance is for small values of $nc$. Sensitivity varies from $57.1\%$ to $88\%$ while specificity from 
$86.33\%$ to $93.84\%$. All the corresponding PTE variants outstand the TE (mean F1-score $48.2\%$) and the PTE (mean F1-score $36.18\%$).
The optimal PTE variants are displayed in Table~\ref{tab:cHenonK10PTEvariants}. Cases 3B and 1B for small $nc$ outperform the other cases. Thus, conditioning on the most correlated variables based on MI seems to be the best approach to improve the performance of PTE. 	
\begin{table}[ht]
\centering
  \caption{Classification measures for optimal connectivity-based PTE variants for S3 in $K=10$ variables, as a mean over all realizations and time series lengths.}
 {\begin{tabular}{ccccc} \hline
 PTE variant & nc    & Sen      & Spec    & F1-score        \\ \hline
 3B              & 2      & 88         & 90.46   & 76         \\
 3B              & 3      & 82.81   & 92.09   & 75.43     \\
 1B              & 2      & 78.62   & 93        & 74.61     \\   
 1B              & 3      & 76.6     & 93.55   & 74.09     \\  
 3B              & 4      & 76.44   & 92.89   & 72.84     \\
 1B              & 4      & 73.21   & 93.78   & 72.24     \\
 3A              & 3      & 79.42   & 90.37   & 71.04     \\
 1A              & 3      & 77        & 90.77   & 70.13   \\ \hline
 \end{tabular}
\label{tab:cHenonK10PTEvariants}}
\end{table}

To further examine the performance of the optimal connectivity-based variant 3B, we display the extracted outcomes for all $nc$ (Table~\ref{tab:cHenonK10_3B_all_nc}).
 As $nc$ increases, sensitivity decreases, varying from $55\%$ to $58.58\%$. Specificity achieves higher scores and has a smaller variation. It increases 
from $nc=2$ up to $nc=5$, where gets its highest value and then decreases again. The overall performance based on F1-score deteriorates as $nc$ grows. 
\begin{table}[ht]
\centering
  \caption{Classification measures for PTE variant 3B for S3 in $K=10$ variables, as a mean over all realizations and time series lengths.}
 {\begin{tabular}{c c c c } \hline
  $nc$ & Sen      & Spec    & F1-score      \\ \hline
  2     & 88        & 90.46   & 76         \\
  3     & 82.81    & 92.09   & 75.43     \\
  4     & 76.44    & 93       & 74.61     \\
  5     & 68.77    & 93.55   & 74.09     \\  
  6     & 62.12   & 92.89   & 72.84      \\
  7     & 58.58   & 93.78   & 72.24      \\   \hline
 \end{tabular}
\label{tab:cHenonK10_3B_all_nc}}
\end{table}
The time series length also affects the exported results (Table~\ref{tab:cHenonK10_3B_all_n}). As $n$ gets larger, sensitivity improves, but specificity worsens. The overall performance of the measure gains strength as $n$ advances. The described course of progress regarding $nc$ and $n$ is common for all PTE variants.
\begin{table}[ht]
\centering
  \caption{Classification measures for PTE variant 3B with $nc=2$, for S3 in $K=10$ variables, as a mean over all realizations.}
 {\begin{tabular}{c c c c } \hline
  $n$       & Sen      & Spec    & F1-score        \\ \hline
  512      & 79.25   & 91.05   & 71.88      \\
  1024    & 89.44   & 90.43   & 76.79     \\
  2048    & 95.31   & 89.91   & 79.33     \\   
  MEAN   & 88        & 90.46   & 76        \\  \hline
 \end{tabular}
\label{tab:cHenonK10_3B_all_n}}
\end{table}

Finally, the mean F1-score of the RF-based variants is $73.84\%$, $75.97\%$ and $46.11\%$ for 5A, 5B and 5C, respectively (Table~\ref{tab:S3K10PTEvariantsRF}). Large time series lengths are required to reach high sensitivity. Specificity is high for 5A and 5B, however both decrease with $n$. Variant 5B slightly outperforms all other cases and 5A follows closely scoring a bit lower than optimal connectivity-based variants. The poor performance of 5C is due to the increased number of conditioning variables considered. This outcome, confirms the necessity of forming a small subset of conditioning variables in order to face the curse of dimensionality. 
\begin{table}[ht]
\centering
\caption{Classification measures for PTE variants based on RF for S3 in $K=10$ variables, as a mean over all realizations for each $n$.}
 {\begin{tabular}{c  ccc  ccc  ccc} \hline
 &\multicolumn{3}{c}{$n=512$}   & \multicolumn{3}{c}{$n=1024$}   & \multicolumn{3}{c}{$n=2048$}   \\ 
 & Sen & Spec  &  F1-score & Sen & Spec &  F1-score  & Sen & Spec &  F1-score    \\  \hline
 5A &  68.63  & 93.51  & 69.17 & 79.31 & 92.78  & 74.71  & 84.13  & 92.86   & 77.64  \\   
 5B & 75.12   & 92.55  & 71.68 & 85.56  & 91.41  & 76.09 & 91.5 & 91.88  & 80.14 \\ 
 5C & 87.25   & 64.49  & 50.01 &  92.19  & 53.74   & 45.6  & 93.5  & 46.91  & 42.71 \\ \hline
\end{tabular}
\label{tab:S3K10PTEvariantsRF}}
\end{table}

\section{Conclusion}
\label{sec:Conclusion}

Real world systems, such as from neuroscience and finance, usually exhibit nonlinear dynamics. Therefore, sophisticated methods are required to reveal the direction of the driving forces among the examined variables, in order to address the limitations of linear tests. Causality analysis provides a variety of methods capable of detecting directional relationships, such as transfer entropy (TE) and partial TE (PTE). This article introduces an ensemble of PTE variants, that aim to overcome the limitations of TE and PTE in case of high dimensional systems. Their advantages are highlighted mainly for high-dimensional systems, where TE and PTE perform poor. 

The connectivity-based PTE variants (1A - 4B) are defined using variable selection in order to reduce the number of conditioning variables and subsequently reduce the dimensionality. The number of conditioning variables $nc$ is defined a'priori in terms of the number of the variables of the simulation system. For the PTE variants using the random forest (RF) method (5A-C), the number of conditioning variables is automatically estimated using the extracted variable importance (VIMP). The proposed PTE variants are examined in a simulation study where different types of simulation systems are considered, i.e. a linear stochastic system (S1), a stochastic system with linear and nonlinear couplings (S2) and a chaotic system with an intermediate coupling strength among the variables (S3). The systems have been selected in order to have different characteristics, i.e. unidirectional and bidirectional couplings, linear and nonlinear couplings and common drivers.  

The TE, although commonly used on a variety of applications, seems to be problematic at cases. First of all, the TE is bivariate and therefore indicates both direct and indirect causal effects. However, the reduced specificity of TE in this simulation study, is only partly due to the detection of indirect causality. The inability of TE to exploit all the available information of the data, results in the identification of spurious directed links. Therefore, this is the main drawback of TE.

The PTE, is the multivariate extension of TE. By definition, it should only identify direct interrelationships. Although effective on low-dimensional systems, turns out to be problematic for high dimensional data sets. The main disadvantage of this approach is that it requires large data sets to achieve a high sensitivity. However, as the dimension of a system increases, it may not be possible to consider an adequate data size to reach a high sensitivity or to balance between sensitivity and specificity, since specificity decreases with the time series length. We should note that spurious causal influences may also appear with PTE, however seems to indicate fewer false links compared to TE in low dimensional systems.

The connectivity-based PTE variants seem to be effective, especially for high-dimensional systems. The time series length and the number of conditioning variables affects their performance, i.e. large $n$ and small $nc$ should be considered. Their performance for S1 and S2 is roughly the same, i.e. they score very close to the optimal PTE and outperform the bivariate TE. Although achieving a high sensitivity, their specificity is not that high, influencing their overall F1-score. Their effectiveness is evident in the chaotic system S3, whereas outperform TE and PTE for both $K=5$ and $K=10$, scoring much higher. Although we do not obtain the same optimal PTE variant consistently for all systems, we end up with an ensemble of conditioning variables that are mostly correlated with the driving or response variable. For S1, the optimal PTE variant is 3A; it is reasonable to come up with a case that considers the linear CC (and not MI), since S1 is linear. For S2 and S3, which are nonlinear systems, we end up with the optimal variants 1B and 3B, i.e. variants that are based on the general correlations of the examined system computed by MI. 

Finally, we introduced the RF-based PTE variants. An additional advantage of these variants stands in the automatic selection of the number of conditioning variables. However, they do not have a consistent performance in the examined systems. Variants 5A and 5B score among the best measures for S1 and S3, while perform really poor for S2. 
Variant 5C is optimal for S1 and S2 but scores last for S3. The poor performance of 5C on high-dimensional systems is due to the fact that 5C, by construction, considers an increased number of conditioning variables compared to 5A and 5B. 

To conclude, all the extracted results highlight the necessity of constructing a reduced subset of conditioning variables in order to face the curse of dimensionality when estimating multivariate causality measures. Although results are displayed for PTE, the suggested variants can be applied to any multivariate causality measure.  
 
\section*{Acknowledgments}
This project has received funding from the Hellenic Foundation for Research and Innovation (HFRI) 
and the General Secretariat for Research and Technology (GSRT), under grant agreement No 794.

\bibliographystyle{unsrt}
\bibliography{PTEvariants}

\end{document}